\documentclass{appolb}%
\usepackage{epsfig}
\usepackage{amsmath}
\usepackage{subfigure}%
\usepackage{amsfonts}%
\usepackage{amssymb}%
\usepackage{graphicx}

\begin{document}

\title{Phenomenological improvement of the linear-$\sigma$ model in the large-$N_{c}$
limit\thanks{Presented by A. Heinz at the Excited QCD Workshop, 20.2.-25.2.2011, in Les Houches
(France)}}
\author{Achim Heinz
\address{Institute for Theoretical Physics, Johann Wolfgang Goethe University,
Max-von-Laue-Str.\ 1, D--60438 Frankfurt am Main, Germany} }
\maketitle

\begin{abstract}
The linear-$\sigma$ model has been widely used to describe the chiral phase
transition. Numerically, the critical temperature $T_{c}$ of the chiral phase
transition is in agreement with other effective theories of QCD. However, in
the large-$N_{c}$ limit $T_{c}$ scales as $\sqrt{N_{c}}$ which is not in line
with the NJL model and with basic expectations of QCD, according to which
$T_{c}$ is --just as the deconfinement phase transition- $N_{c}$-independent.
This mismatch can be corrected by a phenomenologically motivated temperature
dependent parameter.

\end{abstract}




\section{Introduction}

\label{1}The Quantum Chromodynamics (QCD) at finite temperature and finite
density is a central topic in high energy physics. For small temperatures and
densities the quark and gluon degrees of freedom are confined in hadrons. It
is expected that there exists a region in the QCD phase diagram where quarks
and gluons behave like a plasma (deconfinement)
\cite{Cabibbo:1975ig,Rischke:2003mt}. The QCD Lagrangian does not allow to
directly calculate  the confinement/deconfinement phase transition or the
related chiral phase transition. The latter is mathematically well defined in
the limit of zero quark masses, in which the QCD-Lagrangian is invariant under
chiral symmetry transformation and the chiral condensate is an exact order
parameter \cite{Casher:1979vw,Cheng:2006qk,Aoki:2006br}. Two effective models,
the Nambu Jona-Lasino model (NJL)
\cite{Nambu:1961tp,Klevansky:1992qe,Vogl:1991qt,Hatsuda:1994pi} and the linear-$\sigma$ 
model \cite{Gell-Mann:1960ls,Giacosa:2006tf,Parganlija:2010fz}, have
been often used to study the properties of the chiral phase transition. 

Beside these phenomenological approaches to QCD there is also the large-$N_{c}$ 
approximation \cite{'tHooft:1973jz,Witten:1979kh}. The number of color
degrees of freedom in the QCD Lagrangian is three, but a theory with an
infinite large number of color degrees of freedom shows a behavior similar to
the one of a theory with three colors. In the large-$N_{c}$ limit the gauge
symmetry of the QCD is changed from $SU(3)$ to $SU(N_{c}\gg3)$. Enlarging the
number of colors also leads to a modified QCD coupling $g_{QCD}$ in a way that
for $N_{c}\rightarrow\infty$ the product $g_{QCD}^{2}N_{c}$ remains constant.
Quarks and gluons are still present but gluons dominate the behavior of the
theory. For low temperatures there still exists an confined phase, where the
degrees of freedom are mesons and baryons \cite{Bonanno:2011yr}. For high
enough $T$ it is believed that the theory is deconfined. Although also in the
large-$N_{c}$ limit the theory is not solvable, it is significant simpler
because only planar diagram survive.

\section{Linear-$\sigma$ model in the large-$N_{c}$ limit}

\label{3} The linear-$\sigma$ model \cite{Gell-Mann:1960ls,Giacosa:2006tf,Parganlija:2010fz}, is
an effective theory which is able to describe the mass splitting of the pions
and the sigma via spontaneous symmetry breaking. The model is built with terms
which are invariant under chiral symmetry transformation. In the vacuum the
chiral symmetry is spontaneously broken and the pions emerge as Goldstone
bosons. In the original form there is no explicit $N_{c}$ dependency. From
former studies one knows that the quark-antiquark meson masses are independent
of the number of colors, but the coupling of three mesons is suppressed by a
factor of $1/\sqrt{N_{c}}$ and the four mesons coupling by a factor of
$1/N_{c}$ \cite{Witten:1979kh}. These scaling properties can be implemented by
redefining the meson four point interaction $\lambda\rightarrow3\lambda
/4N_{c}$, while the parameter $\mu$ is not affected in the large-$N_{c}$
limit: $\mu\rightarrow\mu$. The Lagrangian of the $\sigma$-model as function
of $N_{c}$ reads:%

\begin{equation}
\mathcal{L}_{\sigma}(N_{c})=\frac{1}{2}(\partial_{\mu}\Phi)^{2}+\frac{1}{2}%
\mu^{2}\Phi^{2}-\frac{\lambda}{4} \frac{3}{N_{c}} \Phi^{4} \text{ ,}
\label{ls1}%
\end{equation}
where the scaler field $\sigma$ and the pseudoscalar pion triplet $\vec{\pi}$
are described by $\Phi^{t} = (\sigma, \vec{\pi})$. For $\mu^{2} > 0$ the
chiral condensate is $\varphi_{0} = \varphi(T=0) = \mu\sqrt{N_{c}/ 3 \lambda} =
\sqrt{N_{c}/3} ~ f_{\pi}$. The only scale in the Lagrangian is $f_{\pi}$, to be
identified with the pion decay constant. Note that the chiral
condensate scales with $\varphi_{0} \propto\sqrt{N_{c}}$. The tree-level
masses are not effected by the $N_{c}$-scaling and read $m_{\sigma}^{2} = 3
\lambda f_{\pi}^{2} - \mu^{2}$, $m_{\pi}^{2} = 0$.

The behavior at finite temperature can be analyzed using the
Cornwall-Jackiw-Tomboulis (CJT) formalism \cite{Cornwall:1974vz}. The gap
equation for the chiral condensate is%

\begin{equation}
0=\varphi(T)\left(  \frac{3}{N_{c}}\lambda\varphi(T)^{2}-\mu^{2}+\frac
{9}{N_{c}}\lambda\int G_{\sigma}+\frac{9}{N_{c}}\lambda\int G_{\pi}\right)
\text{ .} \label{sigma_gap}%
\end{equation}
The full propagators $G_{\sigma}$ and $G_{\pi}$ have the form:%

\begin{equation}
G_{i}=\int_{0}^{\infty}\frac{dk~k^{2}}{2\pi^{2}}\frac{1}{\sqrt{k^{2}
+m_{i}^{2}}} \left[  \exp\left(  \sqrt{k^{2}+m_{i}^{2}} / T\right)  -1 \right]
^{-1} \text{ .}%
\end{equation}
The critical temperature $T_{c}$ is defined as the temperature where the
condensate exactly vanishes $\varphi(T_{c}) = 0$. This leads to the following
$N_{c}$ dependent scaling of the critical temperature:%

\begin{equation}
T_{c}(N_{c})=\sqrt{2}f_{\pi}\sqrt{\frac{N_{c}}{3}}\propto N_{c}^{1/2}\text{
.}\label{tcs}%
\end{equation}
For the case $N_{c}=3$ obtains the known result $T_{c}=\sqrt{2}f_{\pi}$, but
it also clearly emerges that $T_{c}$ increases with $N_{c}$. This means that for
$N_{c}\gg3$ the chiral phase transition will not take place. The result is
general for all hadronic models which do not include color degrees of freedom
or temperature dependent parameters. This result, first noticed in Ref.
\cite{Megias:2004hj}, contradicts the results found in the NJL-model
\cite{Klevansky:1992qe} where the critical temperature $T_c$ remains constant in the
large-$N_{c}$ limit, and with the fact that the related
confinement/deconfinement phase transition is expected to be proportional to
$\Lambda_{QCD},$ which is a large-$N_{c}$ independent quantity.

\section{Phenomenological modification of the linear-$\sigma$ model}

\label{4} In order to solve this discrepancy a phenomenological approach is
proposed. In Refs. \cite{Gasser:1986vb,Leupold:2010zz} it is argued that the
$T^{2}$ scaling of order parameters is general. A phenomenological way to take
this property into account is to make the parameter $\mu^{2}$ temperature
dependent:%

\begin{equation}
\mu^{2}\rightarrow\mu(T)^{2}=\mu^{2}\left(  1-\frac{T^{2}}{T_{0}^{2}}\right)
\text{ .} \label{mod_mu}%
\end{equation}
The parameter $T_{0}$ is a new temperature scale and should be of the
order of $\Lambda_{QCD}$. The Eq. (\ref{mod_mu}) modifies the
gap equation (\ref{sigma_gap}) and leads to a different critical temperature:%

\begin{equation}
T_{c}(N_{c})=T_{d}\left(  1+\frac{T_{d}^{2}}{2f_{\pi}^{2}}\frac{3}{N_{c}%
}\right)  ^{-1/2}\text{ }\propto N_{c}^{0}\text{ .}%
\end{equation}
The critical temperature is constant in the limit $N_{c}\rightarrow\infty$.

Now we turn back to the $N_{c} =3$ and study the case where the explicit
symmetry breaking term $\epsilon\sigma$ is present. The complete Lagrangian
including a second temperature scale reads:%

\begin{equation}
\mathcal{L}_{\sigma}(T_{0})=\frac{1}{2}(\partial_{\mu}\Phi)^{2}+\frac{1}{2}%
\mu^{2}\left(  1-\frac{T^{2}}{T_{0}^{2}}\right)  \Phi^{2}-\frac{\lambda}%
{4}\Phi^{4}+\epsilon\sigma\text{ .}\label{ls2}%
\end{equation}
All parameters are fixed via the masses and the pion decay constant
($\epsilon=f_{\pi}M_{\pi}^{2}$, $\lambda=(M_{\sigma}^{2}-M_{\pi}^{2}%
)/(2f_{\pi}^{2})$, $\mu^{2}=(M_{\sigma}^{2}-3M_{\pi}^{2})/2$). The quantity
$T_{0}$ is set to a value of $T_{0}=0.27\text{~ GeV}$. The vacuum masses
are chosen to be the following: the mass of the $\sigma$-field is $M_{\sigma
}=1.2\text{~ GeV}$ (for the discussion of the value of the $\sigma$ mass in
the vacuum, see Refs \cite{d} and refs therein), of the $\pi$-field is
$M_{\pi}=0.135\text{~GeV}$ and the value for the pion decay constant is
$f_{\pi}=92.4\text{~MeV}$.

\begin{figure}[ptb]
\begin{center}
\includegraphics[scale=0.6] {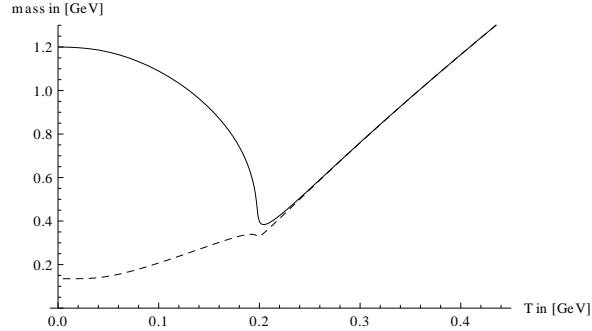}
\end{center}
\caption{Finite temperature behavior of the masses. The dashed line represents
the mass of the pions and the continuous line mass of the $\sigma$. Above an
critical temperature of $T_{c} \approx200 \text{~ MeV}$ the masses become
degenerated.}%
\end{figure}

The finite temperature behavior for the masses, see Fig.\ 1, is similar to the
one with no temperature dependent parameters. Until the critical temperature
$T_{c}$ is reached the temperature dependency of the pion mass an the $\sigma$
mass varies slowly. Close to $T_{c}$ the mass of the $\sigma$ drops and
slightly above $T_{c}$ the mass becomes degenerated with the pion mass. At
high temperature both masses rise linearly.

\begin{figure}[ptb]
\begin{center}
\includegraphics[scale=0.6] {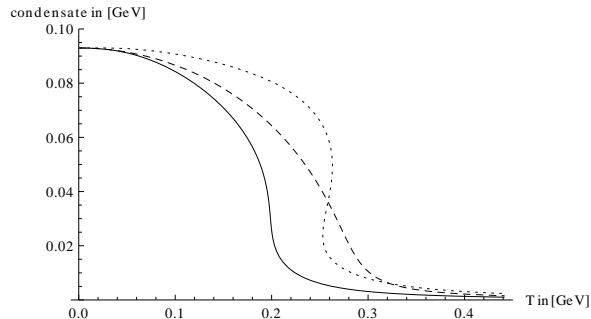}
\end{center}
\caption{The chiral condensate for different number of colors and temperature
scale $T_{0}$. Continuous line: $N_{c} = 3$ and $T_{0} = 270 \text{~ MeV}$,
dotted line: $N_{c} = 3$ and $T_{0} \rightarrow\infty$, dashed line: $N_{c}
\rightarrow\infty$ and $T_{0} = 270 \text{~ MeV}$.}%
\end{figure}

Beside these similarities there are two remarkable properties that differ.
First, the order of the phase transition is changed from a first order to a
crossover phase transition. Second, the critical temperature is lowered to
$T_{c}\approx200\text{~ MeV}$. Both phenomena can be seen in Fig.\ 2., where the
case $N_{c}\rightarrow\infty$ is shown: in this limit the chiral symmetry
is restored through the new temperature scale $T_{0}$ and not via mesonic loops.

\section{Conclusions}

\label{5}In this work the mismatch between the NJL-model and purely
hadronic-models in the large-$N_{c}$ limit has been studied. We have found
that the linear-$\sigma$ model implies a scaling of $T_{c}$ which is at odd
with the NJL model and basic expectations \cite{Casher:1979vw}.

In order to solve this issue we have introduced a phenomenologically
motivated temperature dependent parameter. As a result the critical
temperature remains constant in the large-$N_{c}$ limit. Moreover, for
$N_{c}=3$ the critical temperature is lowered to $T_{c}\approx200$ MeV, a
value which is in line with recent model and lattice results on the chiral
phase transition.

Future studies should go beyond the simple phenomenological Ansatz presented
in this work and include, for instance, the coupling of hadrons to the
Polyakov loop \cite{Rischke:2003mt,Dumitru:2000in}. Preliminary results
\cite{me} show that this approach also leads to the correct large-$N_{c}$
scaling of the critical temperature $T_{c}\sim N_{c}^{0}$.

\bigskip

\textbf{Acknowledgment: }The author thanks F. Giacosa and D. H. Rischke for
cooperation and discussions.


\begin{thebibliography}{99}                                                                                               %




\bibitem {Cabibbo:1975ig}N.~Cabibbo, G.~Parisi,
Phys.\ Lett.\ \textbf{B59}, 67-69 (1975).





\bibitem {Rischke:2003mt}D.~H.~Rischke,
Prog.\ Part.\ Nucl.\ Phys.\ \textbf{52 } (2004) 197-296. [nucl-th/0305030].





\bibitem {Casher:1979vw}A.~Casher,
Phys.\ Lett.\ B \textbf{83}, 395 (1979).






\bibitem {Cheng:2006qk}M.~Cheng, N.~H.~Christ, S.~Datta, J.~van der Heide,
C.~Jung, F.~Karsch, O.~Kaczmarek, E.~Laermann \textit{et al.},
Phys.\ Rev.\ \textbf{D74 } (2006) 054507. [hep-lat/0608013].



\bibitem {Aoki:2006br}Y.~Aoki, Z.~Fodor, S.~D.~Katz, K.~K.~Szabo,
Phys.\ Lett.\ \textbf{B643 } (2006) 46-54. [hep-lat/0609068].





\bibitem {Nambu:1961tp}Y.~Nambu, G.~Jona-Lasinio,
Phys.\ Rev.\ \textbf{122}, 345-358 (1961);
Y.~Nambu, G.~Jona-Lasinio,
Phys.\ Rev.\ \textbf{124}, 246-254 (1961).





\bibitem {Klevansky:1992qe}S.~P.~Klevansky,
Rev.\ Mod.\ Phys.\ \textbf{64}, 649-708 (1992).





\bibitem {Vogl:1991qt}U.~Vogl, W.~Weise,
Prog.\ Part.\ Nucl.\ Phys.\ \textbf{27}, 195-272 (1991).





\bibitem {Hatsuda:1994pi}T.~Hatsuda, T.~Kunihiro,
Phys.\ Rept.\ \textbf{247}, 221-367 (1994). [hep-ph/9401310].





\bibitem {Gell-Mann:1960ls}M. Gell-Mann, M. Levy,
Nuovo\ Cimento,\ \textbf{16}, 705 (1960).
P.~Ko and S.~Rudaz,
Phys.\ Rev.\ D \textbf{50} (1994) 6877.
S.~Gasiorowicz, D.~A.~Geffen,
Rev.\ Mod.\ Phys.\ \textbf{41}, 531-573 (1969). M.~Urban, M.~Buballa,
J.~Wambach,
Nucl.\ Phys.\ \textbf{A697}, 338-371 (2002). [hep-ph/0102260].





\bibitem {Giacosa:2006tf}F.~Giacosa,
Phys.\ Rev.\ \textbf{D75 } (2007) 054007. [hep-ph/0611388].
A.~Heinz, S.~Struber, F.~Giacosa and D.~H.~Rischke,
Phys.\ Rev.\ D \textbf{79} (2009) 037502 [arXiv:0805.1134 [hep-ph]].
A.~H.~Fariborz, R.~Jora and J.~Schechter,
Phys.\ Rev.\ D \textbf{77} (2008) 034006 [arXiv:0707.0843 [hep-ph]].
A.~H.~Fariborz, R.~Jora and J.~Schechter,
Phys.\ Rev.\ D \textbf{72} (2005) 034001 [arXiv:hep-ph/0506170].




\bibitem {Parganlija:2010fz}D.~Parganlija, F.~Giacosa, D.~H.~Rischke,
Phys.\ Rev.\ \textbf{D82}, 054024 (2010). [arXiv:1003.4934 [hep-ph]].
S.~Janowski, D.~Parganlija, F.~Giacosa, D.~H.~Rischke,
[arXiv:1103.3238 [hep-ph]];
S.~Gallas, F.~Giacosa, D.~H.~Rischke,
Phys.\ Rev.\ \textbf{D82 } (2010) 014004. [arXiv:0907.5084 [hep-ph]].





\bibitem {'tHooft:1973jz}G.~'t Hooft,
Nucl.\ Phys.\ \textbf{B72}, 461 (1974).




\bibitem {Witten:1979kh}E.~Witten,
Nucl.\ Phys.\ \textbf{B160}, 57 (1979).





\bibitem {Bonanno:2011yr}Even if baryons exist in the large-$N_{c}$ limit, it
is unclear if nuclear matter binds, see Ref. L.~Bonanno and F.~Giacosa,
arXiv:1102.3367 [hep-ph].




\bibitem {Cornwall:1974vz}J.~M.~Cornwall, R.~Jackiw, E.~Tomboulis,
Phys.\ Rev.\ \textbf{D10}, 2428-2445 (1974).





\bibitem {Megias:2004hj}E.~Megias, E.~Ruiz Arriola, L.~L.~Salcedo,
Phys.\ Rev.\ \textbf{D74}, 065005 (2006). [hep-ph/0412308].





\bibitem {Gasser:1986vb}J.~Gasser, H.~Leutwyler,
Phys.\ Lett.\ \textbf{B184}, 83 (1987).





\bibitem {Leupold:2010zz}S.~Leupold, U.~Mosel,
AIP Conf.\ Proc.\ \textbf{1322}, 64-72 (2010).

\bibitem {d}C.~Amsler and N.~A.~Tornqvist,
Phys.\ Rept.\ \textbf{389}, 61 (2004).
E.~Klempt and A.~Zaitsev,
Phys.\ Rept.\ \textbf{454} (2007) 1.
F.~Giacosa,
Phys.\ Rev.\ D \textbf{80} (2009) 074028.


\bibitem {Dumitru:2000in}A.~Dumitru, R.~D.~Pisarski,
Phys.\ Lett.\ \textbf{B504 } (2001) 282-290. [hep-ph/0010083].



\bibitem {me}A.~Heinz, F.~Giacosa and D.~H.~Rischke, in preparation.


\end{thebibliography}
\end{document}